\documentclass[reprint,superscriptaddress,amsmath,amssymb,aps,prl,tightenlines]{revtex4-2}

\usepackage{hyperref}
\hypersetup{
    colorlinks=true,
    linkcolor=blue,
    citecolor=blue,
    filecolor=magenta,      
    urlcolor=blue,
    pdfpagemode=FullScreen,
    }

\usepackage{gensymb} 
\usepackage{amsmath}
\usepackage{graphicx}
\usepackage{dcolumn}
\usepackage{bm}

\begin{document}

\preprint{APS/123-QED}

\title{Giant Dzyaloshinskii-Moriya interaction and ratchet motion of bimeronic excitations in two-dimensional magnets}

\newcommand{\USTC}{\affiliation{
 International Center for Quantum Design of Functional Materials (ICQD), Hefei National Laboratory for Physical Sciences at Microscale (HFNL), and CAS Center for Excellence in Quantum Information and Quantum Physics, University of Science and Technology of China, Hefei, Anhui 230026, China
}}

\newcommand{\SXNU}{\affiliation{
Key Laboratory of Magnetic Molecules and Magnetic Information Materials of the Ministry of Education and Research Institute of Materials Science, Shanxi Normal University, Linfen 041004, China
}}

\author{Shunhong Zhang}
\altaffiliation{These authors contributed equally to this work.}
\author{Xiaoyin Li}
\altaffiliation{These authors contributed equally to this work.}
\USTC

\author{Huisheng Zhang}
\SXNU

\author{Ping Cui}
\USTC

\author{Xiaohong Xu}
\SXNU

\author{Zhenyu Zhang}
\email{zhangzy@ustc.edu.cn}
\USTC

\date{\today}

\begin{abstract}
Topological magnetic excitations rooted in Dzyaloshinskii-Moriya interaction (DMI) are promising information carriers for next-generation memory and spintronic devices. The recently discovered two-dimensional (2D) magnets provide fertile new platforms for revealing rich physical phenomena with atomic-scale precision, as exemplified by the enhanced DMI associated with an effective electric field that breaks the inversion symmetry. Here we use first-principles calculations to establish a conceptually different compositional engineering approach to induce giant DMI in a representative CrMnI$_6$ 2D magnet, and the underlying mechanism is rooted in the spontaneous inversion symmetry breaking in the bipartite system. Using atomistic magnetics simulations, we further reveal that bimeronic excitations can emerge upon cooling a spin random state or perturbing a ferromagnetic state. Strikingly, the bimeron can exhibit unidirectional soliton-like propagation, and such a ratchet phenomenon is highly desirable for racetrack memories. These findings may prove instrumental in developing novel devices for quantum information based on 2D magnets.
\end{abstract}

\maketitle

The past decades have witnessed the entry and fertilization of the concept of topology in various realms of condensed matter physics, where geometric phase is introduced to effectively characterize the topologically nontrivial electronic and magnetic properties of solids. Different from the electronic band topology~\cite{RN126}, the geometric phase of magnetic structures evolves in real space and manifestation of the corresponding topology is more vivid and intuitive. As a compelling example, the well-known nontrivial magnetic entity of skyrmion cannot be adiabatically deformed into a trivial ferromagnetic state, a distinct and striking property that can be exploited for novel spintronic applications~\cite{RN38}. These promising technological potentials, in turn, have propelled intensive research efforts in searching for more desirable hosting materials, developing better imaging techniques, and achieving more precise control of such topological magnetic quasiparticles ~\cite{RN79,RN1,RN3}.

\begin{figure}[b]
     \centering
     \includegraphics[width=8cm]{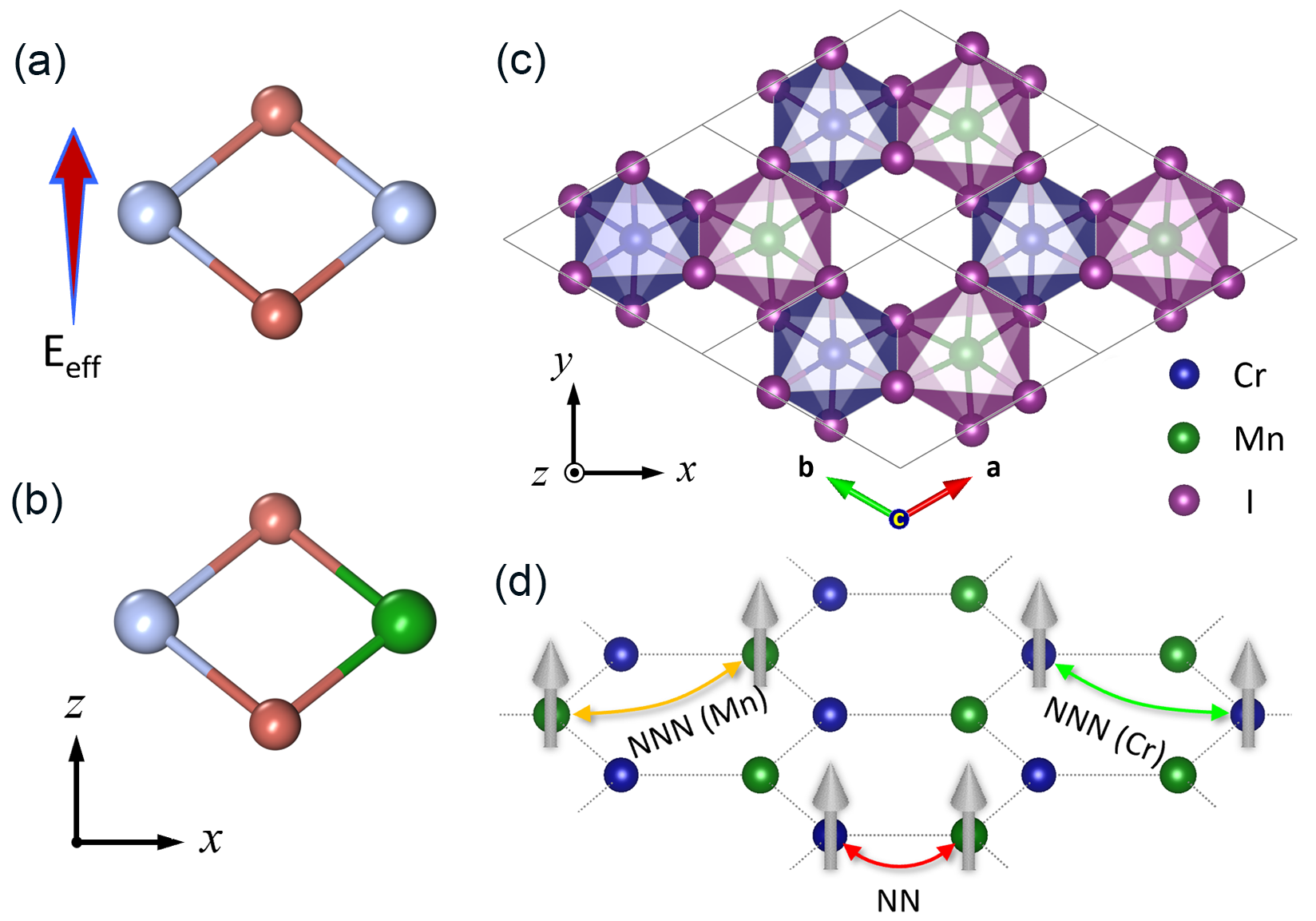}
     \caption{(a) and (b) Schematics of two different approaches to induce DMIs in monolayered magnets (silver/green balls: magnetic cations, orange balls: nonmagnetic anions). (a) The prevailing strategy, relying on an effective vertical electric field $\mathbf{E}_\text{eff}$. (b) Creating an asymmetric bipartite lattice as proposed here. (c) Top view of the atomic structure of monolayered CrMnI$_6$, showing the Cr and Mn ions centered in edge-sharing I$_6$ octahedra. (d) The honeycomb spin lattice with the NN and NNN exchange couplings indicated.}
     \label{fig-1}
 \end{figure}

Although topological magnetic structures encompass abundant forms~\cite{RN1}, their underlying formation mechanisms are predominantly rooted in the same quantum mechanical and relativistic origin, namely, the Dzyaloshinskii-Moriya interaction (DMI)~\cite{RN42,RN119}. Historically, the DMI was introduced to interpret the weak ferromagnetism in prototypical antiferromagnets such as $\alpha$-Fe$_2$O$_3$~\cite{RN42}, but subsequently, it was found that the DMI can also play an essential role in stabilizing the vortex-like configurations in model ferromagnetic systems ~\cite{RN94}. The DMI exchange energy has a chiral form  $\mathbf{D}_{ij}\cdot(\mathbf{S}_i\times\mathbf{S}_j)$, thus favoring perpendicular spin alignment within the plane normal to the DM vector $\mathbf{D}_{ij}$. This antisymmetric exchange coupling of two magnetic moments arises in materials with simultaneous spin-orbit coupling (SOC) and structural inversion symmetry breaking. Recently, layered van der Waals magnets~\cite{RN120}, with CrI$_3$~\cite{RN89} and CrGeTe$_3$~\cite{RN90} as the representatives, have been experimentally thinned to the mono- or few-layer limit while preserving their long-range ferromagnetic (FM) order. As the long-sought members in the portfolio of 2D materials, such 2D magnets provide fertile new platforms for revealing rich physical phenomena with atomic-scale precision, and can also serve as new building blocks in various magneto-devices with higher integration, lower energy consumption, and faster processing~\cite{RN50}. Nevertheless, the DMI is absent in the exchange coupling of two nearest neighbored magnetic sites in pristine CrI$_3$~\cite{RN33}, CrGeTe$_3$~\cite{RN33}, and many other subsequently discovered 2D magnets~\cite{RN70,RN71}, due to their inherent crystalline symmetry~\cite{RN41}. To induce substantially large DMIs, various symmetry-breaking scenarios have been proposed, such as by applying a vertical electric field~\cite{RN73} or invoking Janus structures~\cite{RN9,RN7,RN6} or heterojunctions~\cite{RN45}. In essence, these approaches share the same commonality, namely, by creating an effective electric field perpendicular to the 2D magnets [as sketched in Fig.~\ref{fig-1}(a)], conceptually similar to those extensively studied at interfaces of ferromagnetic and heavy metal films~\cite{RN100,RN28}. The topological magnetic entities associated with such DMIs are typically constrained to be Néel type. Novel centrosymmetry breaking schemes are desired to further enrich and expand the territory of 2D topological magnetism. 
 
This Letter makes multifold contributions. First, we employ a conceptually different design scheme to induce giant DMI in intrinsic yet physically realistic 2D magnets. Here, the underlying physical mechanism does not rely on the establishment of an effective electric field to break the centrosymmetry; instead, the DMI arises from the inherent asymmetry of the magnetic species in the systems [as illustrated in Fig.~\ref{fig-1}(b)]. Specifically, our first-principles calculations reveal that giant Bloch-type DMI exists in a prototypical CrMnI$_6$ monolayer~\cite{RN27}, which is isostructural to CrI$_3$ but with bipartite asymmetry. Salient new magnetic properties are also shown to emerge upon such compositional engineering, including easy in-plane ferromagnetism, as well as anisotropic bilinear and pronounced biquadratic exchange couplings. Secondly, we use atomistic magnetics simulations to demonstrate that paired meronic excitations (bimerons) can be selected upon cooling or controllably excited. Most strikingly, a solitonic bimeron can be driven to move unidirectionally by thermally excited magnons, revealing a ratchet feature in the interior of the in-plane FM racetrack~\cite{RN114,RN112}. This exotic phenomenon, predicted for the first time in 2D magnets, can be further exploited for memory and spintronic device applications. These findings may not only provide instrumental insights to understanding chiral exchange interactions and topological magnetism in the ultrathin regime, but also help to accelerate the applications of 2D magnets in quantum devices such as in racetrack memories~\cite{RN108}.

\textit{Giant DMI in the bipartite CrMnI$_6$ monolayer.}\textemdash The CrMnI$_6$ monolayer shown in Fig.~\ref{fig-1}(c) can be obtained by substituting one sublattice of chromium in the CrI$_3$ monolayer with manganese, and the ordered structure with bipartite spin lattice (Cr$^{3+}$ and Mn$^{3+}$ ions with three and four $\mu_\text{B}$ respectively) as shown in Fig.~\ref{fig-1}(d) is energetically favored~\cite{RN27}. The corresponding Curie temperature was estimated to be enhanced compared to CrI$_3$ within a simplified Heisenberg model, and the system behaves as a magnetic topological insulator exhibiting the quantum anomalous Hall effect~\cite{RN27}. Beyond the nontrivial electronic band topology, the system may possess more direct and intriguing nontrivial magnetic properties in real space. Indeed, the structural symmetry is reduced from D$_{3d}$ to D$_3$ upon substitution. Because of the coexistence of centro-asymmetry and large SOC, the system is expected to harbor pronounced DMI.

To investigate the anisotropy of exchange couplings in the monolayered CrMnI$_6$ system, we first assume an atomistic spin Hamiltonian with bilinear (BL) terms,
\begin{eqnarray}
\label{Eq1}
     H_{BL} = -\sum_{i}{A_{i,zz}S_{iz}^2} - \sum_{i<j}{\mathbf{S}_i\mathbf{J}_{ij}^\text{sym}\mathbf{S}_j} \nonumber \\ -\sum_{i<j}{\mathbf{D}_{ij}\cdot(\mathbf{S}_i\times\mathbf{S}_j)}
    -\mu_\text{B}\sum_{i}{\mathbf{B}\cdot\mathbf{S}_i}.
\end{eqnarray}
The first term represents the single-ion anisotropy arising from atomic SOC, which is reduced to a single scalar due to the three-fold rotational symmetry of the system~\cite{RN29}. The second and third terms correspond to the symmetric and antisymmetric (DMI) parts of the exchange couplings, respectively, with the nearest neighbor (NN) and next nearest neighbor (NNN) terms indicated in Fig.~\ref{fig-1}(d). The last term accounts for the Zeeman energy due to the external magnetic field $\mathbf{B}$.


We perform density functional theory calculation with a mean-field onsite Coulomb repulsion correction for the localized $d$-orbitals (DFT+U), to attain the energetics of the monolayered CrMnI$_6$~\cite{Supp_note,RN131,RN58,RN40,RN54,RN30,RN34,RN77,RN48,RN59,RN60,RN63,RN61,RN47,RN2}. For the magnetic degrees of freedom, we consider four specifically designed non-collinear spin configurations following the powerful four-state method~\cite{RN15,RN29}, from which the exchange parameters can be conveniently extracted with DFT accuracy~\cite{RN20,RN33,RN31,RN32}. The calculated bilinear exchange parameters are summarized in Table S1~\cite{table_note}. One can see that the symmetric exchange couplings are predominantly ferromagnetic, but exhibit highly anisotropic characteristics, as reflected by the distinctly different diagonal values in the $\mathbf{J}^\text{sym}$ tensors and some non-vanishing off-diagonal components. For the antisymmetric part (DMI), more insights can be drawn regarding both directions and magnitudes. For directions, our calculated DM vectors are consistent with the Moriya's rules~\cite{RN41}, namely, the symmetry-forbidden components are numerically negligible as well ($<$0.01 meV). For magnitudes, the $\lvert \mathbf{D} \rvert$/J$_\text{iso}$ ratio~\cite{RN14} has been widely used to evaluate the competition between the DMI and J$_\text{iso}$, which respectively favor canted and collinear alignments of the magnetic moments, with the trace average J$_\text{iso}$ = Tr($\mathbf{J}^\text{sym}$)/3. For the three considered bilinear exchange coupling strengths, the $\lvert \mathbf{D} \rvert$/J$_\text{iso}$ values reach 0.31 (Cr-Mn), 0.44 (Cr-Cr), and 0.82 (Mn-Mn) respectively, the latter two substantially exceeding the corresponding ratios in the prototypical Fe/Ir(111) system~\cite{RN107} and other chiral magnets hosting topological spin textures~\cite{RN7}. These findings indicate that giant chiral magnetic interactions emerge in this system of broken inversion symmetry, a salient feature that is highly desirable for exploring exotic topological magnetism.

\begin{figure}[b]
    \centering
    \includegraphics[width=8.6cm]{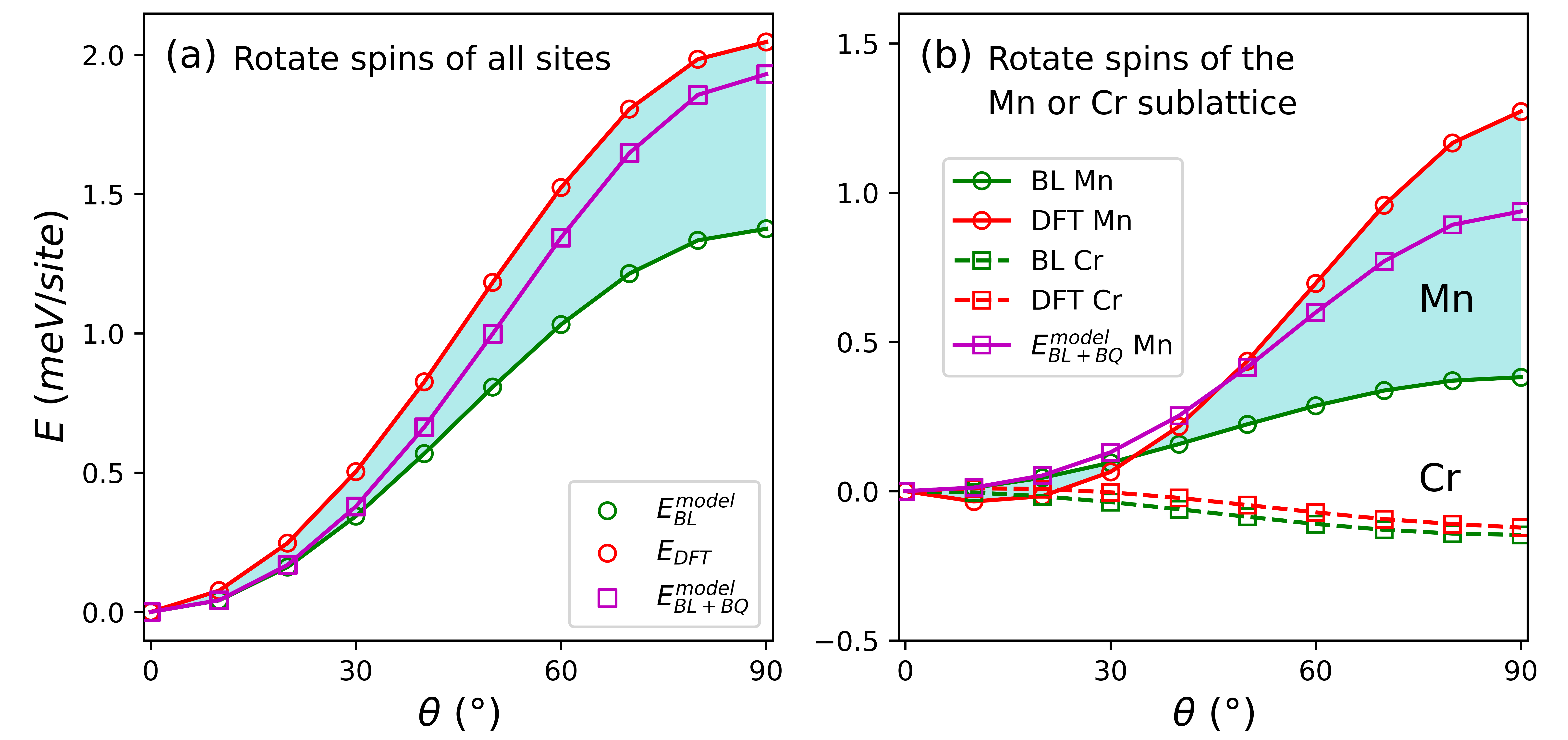}
    \caption{(a) Energy evolution of the FM configuration with rotating magnetization axis. (b) Energy evolution upon rotating spins of the Mn (Cr) sublattice from \textit{x} ($\theta = 0\degree$) to \textit{z} ($\theta = 90\degree$) synchronically while fixing spins of the Cr (Mn) sublattice along \textit{y}. The energy differences between the DFT+U calculations and Eq. (\ref{Eq1}) are shaded in cyan.}
    \label{fig-2}
\end{figure}

\textit{Magnetic anisotropy and biquadratic coupling.} \textemdash Another important physical aspect to investigate is the magnetocrystalline anisotropy energy (MAE), which indicates the easy magnetization orientations. We employ DFT+U calculations to trace the total energy evolution of the FM state upon rotating the spin quantization axis. It is found that the total energy of the system  remains nearly isotropic within the \textit{xy} plane, but is lower than that of the out-of-plane FM state, suggesting that the system favors in-plane magnetization. These features can be correctly predicted by the BL model [Eq. (\ref{Eq1})], indicating its capacity to depict the in-plane physics. We therefore use the in-plane FM configuration as the reference state to calculate the MAE using this model. The overall energy evolution trend can be qualitatively well reproduced [Fig.~\ref{fig-2}(a)], indicating that the MAE mainly stems from competition between bilinear spin-spin interactions. On the other hand, there exists pronounced discrepancy at the quantitative level, which hints on the existence of some complex higher-order spin-spin interactions.

Given the protection of time-reversal symmetry~\cite{RN64}, the likely leading higher-order term is the biquadratic (BQ) coupling~\cite{RN124,RN56}. To reveal the potential existence and microscopic origin of such BQ interactions, we systematically investigate the element-resolved single-ion and pair-resolved ion-ion contributions. The single-ion and NN exchange coupling are both found to be bilinear~\cite{bilinear_note}. Then we study the NNN pairs by rotating the spins of one sublattice synchronically from \textit{x} to \textit{z} while fixing the other sublattice magnetized along $y$~\cite{sublat_note}. By comparing the energy evolutions between the DFT+U and BL model calculations, we see quantitative consistency upon rotating spins of the Cr sublattice. In stark contrast, pronounced differences exist when the spins of the Mn sublattice are rotated [Fig.~\ref{fig-2}(b)], strongly pointing to the likely source of the MAE discrepancy shown in Fig.~\ref{fig-2}(a). Microscopically, this contrast could originate from their different 3$d$-electron fillings in the presence of an octahedral crystal field [Fig.~\ref{fig-1}(c)]. Both ions have three $d$ electrons occupying the low-lying $t_\text{2g}$ orbitals in the spin majority channel, while the Mn$^{3+}$ ions ($d^4$) have an extra $d$ electron populating the higher-energy $e_\text{g}$ orbitals in the same spin channel. The exchange coupling of the $e_\text{g}$ electrons within a Mn-Mn pair in return gives rise to pronounced BQ spin-spin interaction.

We can now further improve the spin Hamiltonian in Eq. (\ref{Eq1}) on solid physics basis. Unlike earlier studied systems with isotropic BQ exchange couplings~\cite{RN124,RN56}, one distinctly new feature of the present 2D system is that the BQ coupling must be highly anisotropic, because it contributes substantially to the MAE. Here we adopt a simple and physically intuitive anisotropic form,
\begin{equation}
\label{Eq2}
H_{BQ} = -\frac{1}{2}\sum_{<<i,j>>}K_{zz}^\text{Mn-Mn}S_{iz}^2S_{jz}^2,
\end{equation}
with $K_{zz}^\text{Mn-Mn}$ being the only coefficient to be determined. With the inclusion of this BQ coupling, the energy evolution trend upon rotating the Mn sublattice spins can be significantly improved, as shown in Fig.~\ref{fig-2}(b). As a crucial crosscheck, it also improves the description of the MAE quantitatively, as confirmed in Fig.~\ref{fig-2}(a). The optimally compromised value that can simultaneously improve the quantitative accuracies in both Fig.~\ref{fig-2}(a) and (b) is -0.023 meV, with the negative sign indicating a BQ contribution to in-plane magnetization. The coefficient seems to be weak at the first glance, but given the large magnetic moments of Mn (S=2), the BL [J$_\text{iso}$S$^2$ = 1.4 meV] and BQ [$K_{zz}^\text{Mn-Mn}$S$^4$ = -0.37 meV] coupling strengths between NN Mn-Mn pairs are readily comparable.

\begin{figure}[b]
    \centering
    \includegraphics[width=8.5cm]{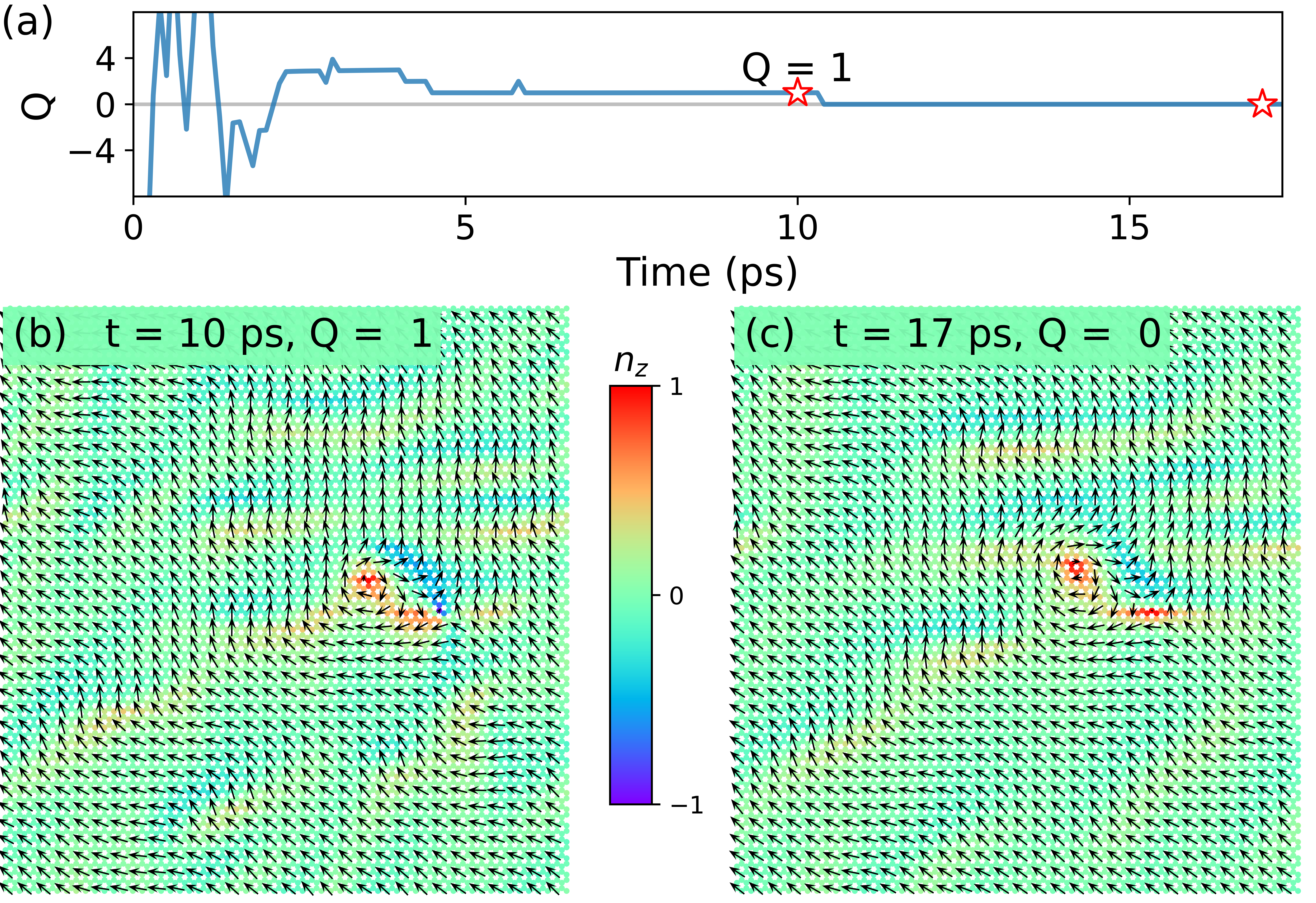}
    \caption{(a) Evolution of the topological charge in an LLG simulation starting from a randomly magnetized configuration. (b) and (c) Snapshot configurations. The in-plane and out-of-plane spin components are shown by the arrows and a rainbow color mapping, as also adopted in Figs. 4.}
    \label{fig-3}
\end{figure}

\textit{Bimeronic topological excitations.}\textemdash The BL-BQ spin Hamiltonian combining Eqs. (\ref{Eq1}) and (\ref{Eq2}) now provides an accurate description of the bipartite spin lattice of monolayered CrMnI$_6$. To explore potential chiral magnetic structures, we perform atomistic magnetics simulations by numerically integrating the Landau-Lifshitz-Gilbert (LLG) equation~\cite{RN48}, which can depict the precession and damping of localized spins in the presence of exchange and external magnetic fields~\cite{LLG_note}.
We start the LLG simulations from a randomly magnetized configuration within a 60×36 supercell which mimics the paramagnetic state. The damping term in the LLG equation dissipates energy of the spin system and drives it to lower energy state, effectively modeling a cooling process~\cite{RN46}. To trace the real space geometric phase of the dynamic spin structure, we calculate the topological charge Q~\cite{RN61} evolution during the simulation. As shown in Fig.~\ref{fig-3}(a), Q fluctuates drastically during the initial transient period of the simulation, and then converges to lower values, indicating the emergence of topological chiral spin textures upon cooling. To directly visualize the whirling of the spin vectors, selective snapshots at $t$ = 10 and 17 ps are plotted. The spin structures clearly show the existence of vortex- or antivortex-like local excitations in the cycloidal background, with different topological charges. Given that the material strongly favors in-plane magnetization, we assign these nontrivial magnetic structures to stem from merons~\cite{bimeron_note} instead of skyrmions that require out-of-plane magnetization~\cite{RN1}. We also note that the pairing of two meronic excitations (also called bimerons) illustrated in Fig.~\ref{fig-3}(b, c), involving one vortex plus one antivortex, fulfills the generic vorticity conservation condition, which is qualitatively analogous to the topological defects of a 2D system below the critical temperature of the Berezinskii–Kosterlitz–Thouless transition~\cite{RN99}. Such bimeronic excitations have been predicted or observed in other easy in-plane magnets~\cite{RN44,RN46,RN32,RN69,RN66}. Similar nontrivial chiral magnetic structures are observed in several independent simulations from different paramagnetic initial states, further confirming that such bimeronic low-energy excitations are the intrinsic property of the monolayered CrMnI$_6$ governed by its giant DMI and in-plane magnetization.
However, such bimeronic quasiparticles selected upon cooling may still possess stochastic uncertainty with regard to their topological charge, size, shape, position, and lifetime when connecting with practical spintronic applications. To overcome such limitations, it is necessary to explore approaches of writing and deleting an individual bimeron in a more controlled fashion~\cite{RN85,RN69}. Our additional comprehensive simulations~\cite{excitation_note} show that an isolated bimeron can be excited from the in-plane FM background with the aid of a pulse-like, confined vertical magnetic field , and can also be conveniently erased by applying a small in-plane magnetic field. Such operating conditions are physically clearly realistic~\cite{RN85}, and the topological excitations with well-defined spatial locations and soliton-like features may exhibit exotic bimeronic dynamics in racetrack geometry.

\begin{figure}[b]
    \centering
    \includegraphics[width=8cm]{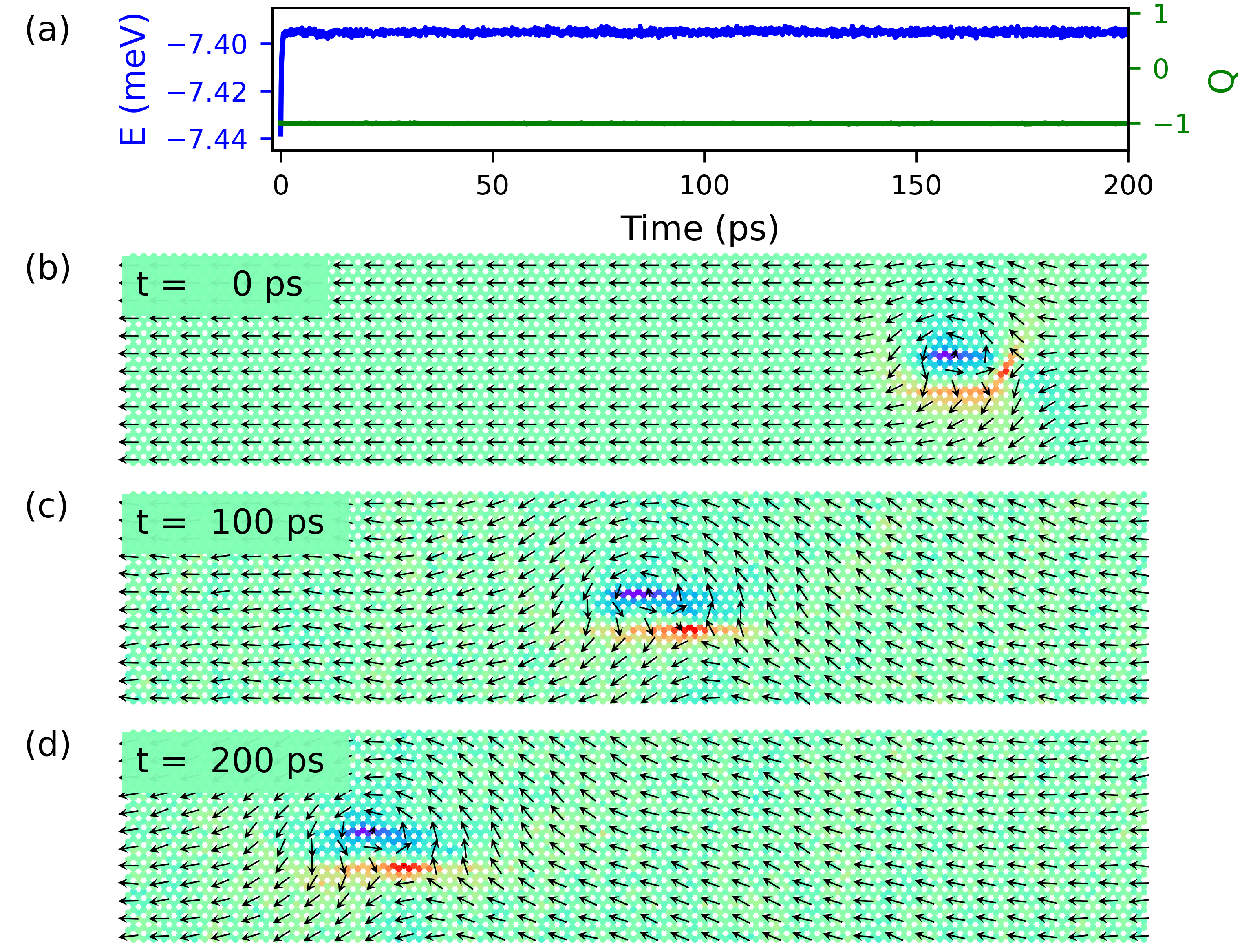}
    \caption{Thermally-driven ratchet motion of a bimeron in racetrack geometry. (a) Evolutions of the total energy/site (left axis) and topological charge (right axis) during the LLG simulations. (b)-(d) Snapshots of the spin configurations.}
    \label{fig-4}
\end{figure}

\textit{Thermally-driven ratchet motion of bimeron.}\textemdash To study the dynamics of bimeron quasiparticles, we create a 100 × 36 supercell in a FM racetrack geometry, and imprint a single Bloch-type bimeron with LLG relaxed configuration. Afterwards, a mild homogeneous thermal field (0.5 K) is additionally exerted to the system. The evolutions of the corresponding spin energy and topological charge are traced, as shown in Fig.~\ref{fig-4}(a). The spin energy increases by $\sim$0.04 meV/site in a few picoseconds, and then fluctuates within a very small energy range, implying that the system has been driven to a steady state. The topological charge conserves the value of -1 throughout the simulation period of 200 ps, implying that the profile of the bimeron is robust and no extra topological excitations are triggered by the thermal field.

To further track the spatial variations of the bimeron, we selectively plot its snapshots in Fig.~\ref{fig-4}(b)-(d). Strikingly, the bimeron not only preserves its original geometry with the mutually bound vortex and antivortex, but its center of mass also exhibits a unidirectional trajectory of motion parallel to the direction of the magnetization, effectively forming a magnetic ratchet~\cite{ratchet_note}. Such intriguing ratchet effects have been predicted or observed in other types of chiral magnetic structures such as domain walls~\cite{RN111,RN109} and skyrmions~\cite{RN112,RN114} within magnetic thin films, but so far has not been reported in any 2D magnet. In particular, the present demonstration also amounts to the first prediction of a ratchet effect surrounding merons. This exotic dynamic behavior can be qualitatively understood from two aspects. The first is about the likely driving force: because no external charge or spin current is applied to induce a spin transfer torque, it is plausible that the thermally excited collective modes in the form of magnons could emerge to provide the underlying impetus propelling the bimeron to propagate like a massive quasiparticle~\cite{magnon_note}. The second is about the unique directionality, whose underlying mechanism could be attributed to the concerted effects of asymmetry as encoded in both the FM configuration and spin Hamiltonian. The former introduces strong anisotropy into the energy landscape, effectively confining the bimeron motion to 1D~\cite{barrier_note}. The latter, featured by sublattice-imbalanced DMIs, cause further differences in the two directions of motion along a 1D channel. The asymmetric dynamics, along with the robust geometry and topological charge, characterizes such bimeronic excitations as reliable information carriers in racetrack applications.

\textit{Discussions.}\textemdash As a practical recipe to induce/enhance chiral spin-spin interactions and stabilize topological magnetic configurations, compositional engineering has been extensively investigated in bulk or thin film materials~\cite{RN97,RN104}. In this Letter, we have explicitly extended the eligibility and feasibility of this design scheme to 2D magnets, which possess inherent advantage in more delicate spin manipulation. Motivated by this proposal, our ongoing experimental efforts have preliminarily shown that layered (Cr, Mn)I$_3$ films can indeed be fabricated via chemical vapor growth, and such samples exhibit novel magnetic properties~\cite{CVD_note}. Alternatively, molecular-beam epitaxy and in-situ doping ~\cite{RN125,RN117} are also powerful techniques for compositional engineering of 2D magnets. Unlike previously studied spin systems~\cite{RN88,RN87,RN122}, the bimeron unidirectional transport predicted here does not rely on specifically designed odd geometry or edges of the materials. The required in-plane FM ground state and sublattice-asymmetric DMI, are both intrinsic properties of the bipartite system, making direct experimental validations more feasible with simpler setups.

\textit{Conclusions.}\textemdash Based on symmetry considerations and first-principles calculations, we have established a powerful compositional engineering scheme to induce pronounced DMI in 2D magnets. Staring from the CrI$_3$ monolayer, the introduction of a second magnetic species Mn can effectively tune the Ising-type ferromagnetism of the parent compound, giving rise to giant DMI and highly anisotropic biquadratic exchange interactions. The bipartite CrMnI$_6$ monolayer has been further shown to exhibit nontrivial magnetic properties featured by nanometric bimeronic topological excitations. Most strikingly, the solitonic bimeron can achieve thermally-driven unidirectional motion, which can be further exploited for racetrack applications. Collectively, these central findings characterize the representative bipartite CrMnI$_6$ monolayer as a new and versatile member in the family of 2D magnets both for intriguing topological magnetism and for conceptually new spintronic devices.

This work was supported by the National Natural Science Foundation of China (Grant Nos. 11904350, 11634011, 11974323, 11722435, and 11804210), the National Key R\&D Program of China (Grant No. 2017YFA0303500), the Strategic Priority Research Program of Chinese Academy of Sciences (Grant No. XDB30000000), the Anhui Initiative in Quantum Information Technologies (Grant No. AHY170000), and Anhui Provincial Natural Science Foundation (Grant. No. 2008085QA30). S. Z acknowledges Dr. C. S. Xu for kind help in the four-state method related calculations.

\bibliography{apssamp}

\begin{thebibliography}{77}
\bibitem [{sub()}]{sublat_note}%
  \BibitemOpen
  \href@noop {} {}\bibinfo {note} {The NNN exchange couplings are in essence
  the NN Cr-Cr or Mn-Mn exchange couplings within the respective sublattices.
  This rotation operation is valid as long as the NNN Cr-Cr or Mn-Mn
  interactions is much weaker than their respective NN
  interactions.}\BibitemShut {Stop}%
\bibitem [{bim()}]{bimeron_note}%
  \BibitemOpen
  \href@noop {} {}\bibinfo {note} {Specifically, we may have merons (Q = -1/2)
  and anti-merons (Q = 1/2), as well as their various paired forms (including
  bimerons with Q = -1, anti-bimerons with Q = 1, and meron plus anti-meron
  pairs with Q = 0). In the rest of this Letter, all the different meron pairs
  are collectively called bimerons, but distinguished with their topological
  charges [e. g. Q = 1 for the bimeron in Fig. 3(b) and Q = 0 for the one in
  Fig. 3(c)].}\BibitemShut {Stop}%
\bibitem [{mag()}]{magnon_note}%
  \BibitemOpen
  \href@noop {} {}\bibinfo {note} {A comprehensive and definitive study of the
  magnon excitations and their delicate interactions with the bimeron is itself
  highly appealing in its own right, and will be the subjects of forth-coming
  investigations}\BibitemShut {NoStop}%
\end{thebibliography}

\end{document}